\documentclass[sigconf, screen]{acmart}
\usepackage{hyperref}
\usepackage{xcolor}
\usepackage{algorithm}
\usepackage{algpseudocode}
\usepackage{pifont}
\usepackage{tikz}
\newcommand*\circled[1]{\tikz[baseline=(char.base)]{
            \node[shape=circle,fill,inner sep=0.5pt] (char) {\textcolor{white}{#1}};}}

\AtBeginDocument{%
  }

\setcopyright{none} 
\copyrightyear{2026}
\acmYear{2026}
\acmDOI{XXXXXXX.XXXXXXX}

\acmConference[MICRO 2026]{The 58th IEEE/ACM International Symposium on Microarchitecture}{October 31--November 04, 2026}{Athens, Greece}

\acmISBN{978-X-XXXX-XXXX-X/XX/XX}

\begin{document}

\title{\vspace{-20pt}ACRONYM: Accelerated Approximate Nearest Neighbor Search in Memory for Dynamic Vector Databases}

\author{Md Mizanur Rahaman Nayan}
\affiliation{
  \institution{Georgia Institute of Technology}
  \country{}
}
\email{mnayan6@gatech.edu}

\author{Tianqi Zhang}
\affiliation{
  \institution{University of California San Diego}
    \country{}
}
\email{tiz014@ucsd.edu}

\author{Flavio Ponzina}
\affiliation{
  \institution{University of California San Diego}
    \country{}
}
\email{fponzina@ucsd.edu}


\author{Tajana Rosing}
\affiliation{
  \institution{University of California San Diego}
    \country{}
}
\email{tajana@ucsd.edu}

\author{Azad J. Naeemi}
\affiliation{
  \institution{Georgia Institute of Technology}
    \country{}
}
\email{azad@gatech.edu}

\begin{abstract}
Vector database search with frequent updates is increasingly critical in applications such as retrieval augmented generation, recommendation systems, and large-scale embedding retrieval. Existing solutions, such as graph-based and partition-based approximate nearest neighbor search (ANNS), suffer from frequent index rebuilding due to data distribution–dependent indexing that impacts continuous deployment and causes long rebuilding latency. This paper proposes an algorithm-hardware co-designed platform, ACRONYM, that addresses key problems with state of the art database search. Algorithmically, it leverages efficient encoding independent of data distribution and Hamming-distance based search for efficient hardware acceleration. Architecturally, we propose CAM-based in-memory parallel distance computation followed by time multiplexed approximated top-k selection to enable the exhaustive search. We propose two-stage search that includes coarse search followed by binary refinement to achieve high recall in CAM based search which is heavily limited to small vector dimension due to capacity and wordline parasitic. ACRONYM supports continuous update without stalling and integrates novel XOR-and-Accumulate (XAC) based systolic-array encoder for efficient on chip encoding during search. Across million-scale datasets, while serving dynamic database ACRONYM achieves $>90\%$ recall at a throughput of $\sim 8\times10^6$ queries per second, with a memory footprint of only $32\mathrm{MB}$ and an average energy consumption of $2.56\mu\mathrm{J}$ per query, speedup over HNSW (CPU) of about $400\times$ and FAISS-IVF (GPU) of about $80\times$.

\end{abstract}


\maketitle

\section{Introduction}

In the era of Large Language Model (LLM)-powered Retrieval-Augmented Generation (RAG), recommendation systems, and data mining, the ability to efficiently search massive high-dimensional databases has become a cornerstone of modern computing \cite{chen2022annsrec, johnson2019billion, lewis2020retrieval}. As data volumes in both public and private repositories continue to grow exponentially, the demand for high-throughput, low-latency similarity search has never been greater.

Among existing approaches, graph-based and partition-based methods dominate due to their superior recall–latency trade-offs at scale\cite{aumuller2020ann, jegou2010product, johnson2019billion}. Graph-based approaches, such as HNSW and DiskANN \cite{jayaram2019diskann, malkov2018efficient}, construct a navigable proximity graph offline, which enables efficient traversal of high-dimensional vector spaces during query time. These methods achieve strong recall while maintaining high query throughput. Partition-based approaches, on the other hand, divide the search space into clusters during an offline phase and restrict the search to a subset of relevant regions at query time\cite{jegou2010product, johnson2019billion, lee2022anna}. This reduces the effective search space and improves efficiency.

However, both classes of methods rely on an initial estimate of the data distribution—either in the form of graph connectivity or clustering centroids—which enables fast navigation and accurate search. This same reliance becomes a limitation in dynamic environments where data is continuously inserted or deleted. As the underlying data distribution evolves, the precomputed graph structure \cite{liu2025wolverine, singh2021freshdiskann, yamashita2025should} or partitioning \cite{mohoney2024incremental} becomes suboptimal, leading to poor performance.

To mitigate this issue, prior work has explored incremental graph updates and adaptive partitioning strategies\cite{zhang2025cleann, mohoney2024incremental}. However, these approaches only partially alleviate the problem as they still dependent on data distribution and often involve graph/partition structural changes, and hence periodic full reconstruction of the index is still required to maintain optimal performance\cite{li2025scalable}.  A similar limitation exists in product quantization (PQ)-based methods, where vectors are compressed using codebooks trained on the initial dataset\cite{jegou2010product}. When the data distribution shifts, retraining the codebooks becomes necessary, leading to performance degradation and system downtime\cite{xu2018online}.

Most recent works on hardware acceleration of ANNS focus primarily on improving throughput and energy efficiency, without addressing the challenges posed by dynamic vector databases e.g., degradation of recall over update cycle and unable to support base vector update while serving\cite{liu2025seim, lee2022anna, yuan2025fanns, wang2024ndsearch, jang2025accelerating,  quinn2025accelerating}. As a result, the recall degradation during serving dynamic DB and stall during update exists in the accelerated systems.

Hashing-based approaches avoid dependence on initial data distribution and are inherently more robust to dynamic updates. However, they have two key limitations: \circled{1} poor scalability on conventional CPU/GPU architectures due to expensive Hamming distance computation and random memory access, \circled{2} reduced recall when using compact hash codes\cite{wang2017survey}. An ideal solution would combine efficient code generation for competent recall with hardware support for massively parallel distance computation, achieving near-constant-time search complexity while minimizing data movement.

This work introduces ACRONYM, a hardware–software co-designed platform for ANNS that enables robust performance under dynamic data updates while maintaining competitive recall–throughput trade-offs to overcome performance degradation of existing approaches with data distribution dependency during dynamic update and hardware implementation challenges of data distribution agnostic hashing-based approach.

Algorithmically, it leverages efficient encoding independent of data distribution and Hamming-distance based search for efficient hardware acceleration. Architecturally, to enable the exhaustive search, demanded by the algorithm, we propose CAM-based in-memory parallel distance computation followed by time multiplexed approximated top-k selection. Top-k selection approach eliminates the requirement for expensive ADC and digtal module for top-k selection out of million scale dataset. To achieve high recall in CAM based search which is heavily limited to small vector dimension due to capacity and wordline(WL) parasitic, we propose two-stage search that includes coarse search followed by binary refinement. Additionally, we introduce XOR-and-Accumulate(XAC) based systolic-array encoder for efficient encoding during search. 

In this co-designed framework, data transfer is limited to a small set of refinement candidate codes between memory and the CAM unit. This significantly improves throughput compared to conventional CPU/GPU-based systems, which suffer from off-chip communication. We propose tightly coupled grouped data address encoding and decoding to further reduce memory access overhead. Furthermore, we demonstrate that the two-stage search strategy not only achieves high recall but also alleviates the challenges of mapping large-scale vector datasets onto CAM arrays under device and circuit level constraints.

In summary, ACRONYM enables consistent and continuous high-throughput performance for dynamic vector databases through the following key innovations:

\begin{itemize}
\item \textbf{Algorithm.} Hardware mapping aware two-stage search framework, combining projection-based encoding with coarse search and refinement to achieve high recall in dynamic database where data are often from out of order distribution.
\item \textbf{Architecture.} Fully static coarse-search pipeline mapped onto CAM, eliminating data movement during the coarse phase and minimizing data transfers during refinement. 
\item \textbf{Architecture.} CAM-based fully in-memory distance computation with approximate top-k selection avoiding expensive distance computation and sorting logic.
\item \textbf{Architecture.} XAC unit based systolic-array encoder for efficient encoding during search that replaces multipliers and improves data reuse.
\item \textbf{Design Space Exploration.} A comprehensive scalability analysis across CAM technologies, identifying optimal design trade-offs under device and system constraints.
\end{itemize}

\begin{figure}
    \centering
    \includegraphics[width=1.1\linewidth]{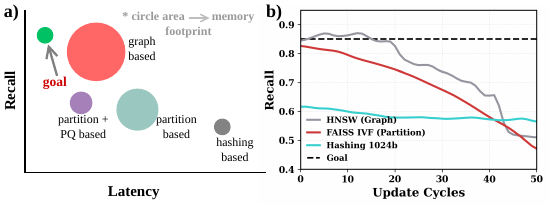}
    \vspace{-20pt}
    \caption{Trade-off in different ANNS method. a) Recall vs Latency with memory size dictated by circle area. b) Recall vs Update cycle that dictates dynamic environment scenario where data distribution shift degrades performance. }
    \vspace{-15pt}
    \label{fig:acro_motiv}
\end{figure}


\section{Background and Motivation}

  



ANNS aims to efficiently retrieve the top-k vectors closest to a query $q \in \mathbb{R}^d$ from a large-scale dataset of size $N$, trading off exactness for substantially higher throughput. Search quality is typically measured using recall@$K$, while performance is quantified in queries per second (QPS). For large-scale deployments and edge applications, memory footprint is an additional critical consideration. An ideal ANNS system achieves high recall under strict latency constraints while maintaining low memory overhead.
\subsection{Existing ANNS Methods and Trade-off}

Fig.\ref{fig:acro_motiv}(a) illustrates the trade-off in recall vs latency between different methods with memory footprint denoted by circle area. Graph based approach offers high recall vs latency but requires large memory to store graph and vectors. Partition based offers medium to high recall at moderate latency due to large number of distance computation involved. Product quantization based offers very small memory footprint due to compressed data storage but trade-offs with recall. The Hashing methods have small memory footprint but suffer from long latency at scale due to exhaustive distance computation and memory communication. 

\subsection{Challenges: ANNS in Dynamic Vector DB}
\begin{figure}
    \centering
    \vspace{-5pt}
    \includegraphics[width=0.8\linewidth]{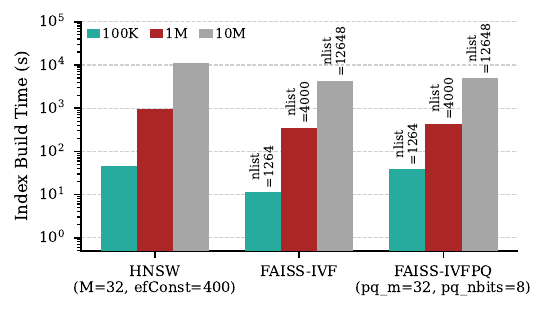}
    \vspace{-15pt}
    \caption{Index building time for different ANNS method.}
    \vspace{-15pt}
    \label{fig:index_building_time_comp}
\end{figure}

Modern vector databases are inherently dynamic, with frequent insertions and deletions. This introduces several fundamental challenges because most ANNS methods rely on an initial estimate of the data distribution\cite{baranchuk2023dedrift}. With new data insertion from different distribution, precomputed graph structures or cluster centroids become suboptimal. Recent work has shown that such drift leads to degraded recall and search efficiency over time \cite{singh2021freshdiskann, liu2025wolverine}. To highlight how performance drifts over updates, we performed basic update with 5\% data random deletion and insertion at each cycle for all the approach.  Fig.\ref{fig:acro_motiv}(b) illustrate the recall of the ANNS methods at different update cycle and shows that they degrade quickly over update cycle, necessarily require graph re-builiding or re-partioning. To maintain performance in dynamic scenario, developing policy regarding graph or centroids updates are under active research\cite{baranchuk2023dedrift, aden2025quantization, zhong2025lsm, zhang2025cleann}. However, graph-based and partition-based methods still require more extensive search which increase latency and impact throughput. Another solution is periodic rebuilding or retraining. These are computationally expensive and can introduce system downtime or significant background overhead \cite{jayaram2019diskann, yamashita2025should}. To highlight the severity, Fig.\ref{fig:index_building_time_comp} illustrates the index building and repartitioning time across ANNS methods in the system mentioned in the experiment section.


\subsection{Challenges of Hashing based technique}
From Fig.\ref{fig:acro_motiv}(b) note that naive hashing based method demonstrates superior immunity against update cycles due to data-distribution-agnostic nature. In principal, hashing-based techniques map high precision vectors into compact binary codes, enabling efficient Hamming distance computation and low memory footprint~\cite{wang2017survey,ong2016improved,  gionis1999similarity}. Data-distribution-agnostic hashing via random projections with binarization provides robustness to distribution drift and offers updates without retraining~\cite{achlioptas2003database}. It also improves over naive hashing approach in search accuracy\cite{norouzi2013fast}. However, evaluating these codes at scale on conventional von-Neumann architectures is fundamentally memory-bound\cite{lee2022anna, norouzi2013fast}. This scenario becomes worse with large hashing code which is crucial for high recall\cite{norouzi2013fast}. Because the number of distance computations scales linearly with dataset size, these workloads rapidly saturate on-chip buffers and cache hierarchies, resulting in severe off-chip memory-to-CPU communication overhead\cite{ibrahim2024efficient}. This fundamental mismatch between hashing workload requirements and von-Neumann bottleneck motivates custom algorithm-hardware co-designed platform. By integrating compute-in-memory (CIM), exploiting bit-level parallelism, and efficiently managing irregular access patterns, such architectures can enable scalable, high-throughput, dynamic vector database search.
\begin{table}[htb]
\centering
\vspace{-5pt}
\caption{Comparison of CAM Technologies}
\vspace{-5pt}
\label{tab:memory_comparison}
\resizebox{\columnwidth}{!}{%
\begin{tabular}{@{}lccccc@{}}
\toprule
\textbf{Metric} & \textbf{CMOS} & \textbf{FeFET} & \textbf{ReRAM} & \textbf{SOT-MRAM} & \textbf{PCM} \\ \midrule
Reference & \cite{9373942} & \cite{9372119} & \cite{10745585} & \cite{11278694} & \cite{li20131} \\
Cell type & 10T & 2 FeFET & 2T-2R & 3T-2MTJ & 2T-2R \\
Node (nm) & 28 & 45 & 90 & 7 & 45 \\
Cell area ($\mu$m$^2$) & 2.66 & 0.15 & 0.41 & 0.08 & 0.41 \\
Search energy (fJ/bit) & 1.02 & 0.35 & 0.55 & 0.714 & 0.64 \\
Endurance\cite{haensch2023compute, yu2022semiconductor} & $>10^{16}$ & $10^9$ & $10^5$ & $10^{12}$ & $10^7$ \\ \bottomrule
\end{tabular}%
}
\vspace{-10pt}
\end{table}

\begin{figure}
    \centering
    \includegraphics[width=\linewidth]{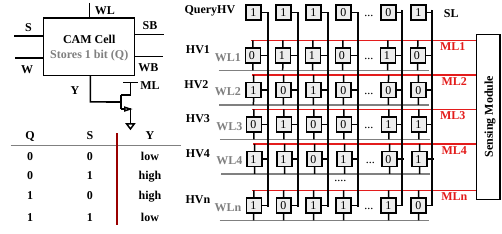}
    \vspace{-20pt}
    \caption{Content addressable memory cell, functionality and mapping of similarity search.}
    \vspace{-15pt}
    \label{fig:cam_search}
\end{figure}

\subsection{In Memory Computing with Content Addressable Memory and Challenges}

To mitigate memory wall, In-Memory Computing (IMC) has emerged as a highly efficient paradigm\cite{roy2025breaking}. By performing computations directly within the memory arrays where the data resides, IMC drastically reduces data movement, offering immense, highly parallelized throughput\cite{roy2025breaking,liu2025seim}. Performing search operation in memory requires distance computation between query vector and candidate vectors followed by finding the most similar candidate based on the computed distances. For Hamming distance computation in memory, content addressable memory has achieved great attention through promising results\cite{sun2023full, 11278694,ni2019ferroelectric}. Fig.~\ref{fig:cam_search} illustrates a CAM cell and its mapping to a Hamming distance based similarity search within an array. In the setup each CAM cell operates as an XOR-like comparator: a mismatch between the stored bit (Q) and the search bit (S) activates a discharge path on the matchline (ML) by a voltage drop at the gate (Y) of driving transistor. Cells connected along a common ML collectively enable the measurement of the Hamming distance in the form of time to discharge\cite{ni2024tap} or current required to sustain the ML voltage\cite{imani2017exploring} between the query vector applied at the S nodes and the stored base vector encoded across the memory cells. Since each row stores a different base vector, multiple base vectors can be compared with a query simultaneously. As a result, distance computation for all stored vectors can be performed in parallel on their respective MLs, achieving constant-time, O(1), distance computation in memory. The resulting ML signals, which encode the distances, are subsequently processed by a comparator block to identify the top similar stored vectors. Although CAM mainly developed for exact matching to be used in network IP finding, they are currently used for most similar item search\cite{nayan2025hydra, imani2017exploring, yu2025hdanns}. However, the application comes with the following additional challenges: \\
\textbf{Overhead for Top-k Selection.} Finding most similar items requires expensive comparator block in terms  of area, energy and delay\cite{nayan2025hydra, liu2022cosime}. Usually hierarchical tree like structures are used to find the most similar candidate and thus responsible for linear delay w.r.t number of stage\cite{imani2017exploring, liu2022cosime}. Additionally, number of leaf blocks in the tree increases linearly and thus gets prohibitive at scale of million scale similarity search. Moreover, for finding top-k items needed by ANNS instead of most similar candidate, the tree like structure get even larger and thus alternative approach could be to use near-memory peripheral circuitry—such as multi-bit Analog-to-Digital Converters (ADCs). Subsequently, digital top-k sorting blocks must be implemented as peripherals to rank the outputs and return the nearest neighbors. This approach also demands large number of ADC causing large area, energy overhead or delay due to limited number of ADCs being shared across huge number of ML along with digital block overhead. To this end, ACRONYM offers time multiplexed approximate top-k selection approach which eliminates the need of tree like structure through time based latching.\\
\begin{figure*}
    \centering
    \vspace{-15pt}
    \includegraphics[width=\textwidth]{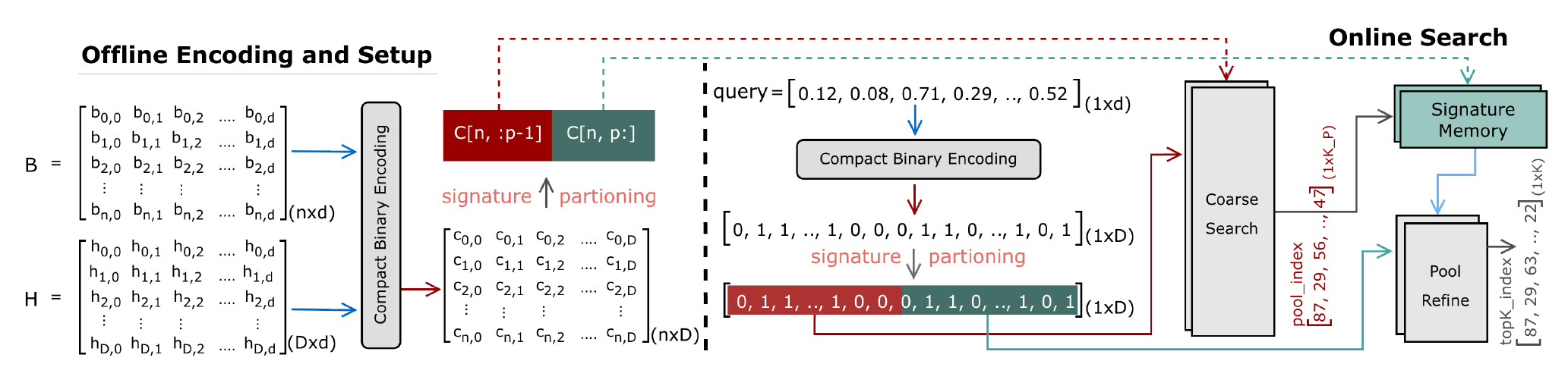}
    \vspace{-25pt}
    \caption{ACRONYM algorithmic flow. In offline base vectors are encoded and partitioned into two code. One partition is stored in coarse CAM block for coarse search and another one in signature memory. During search, query vector is encoded using the same projection matrix used in offline to generate binary code for coarse search and then refinement. Refinement is performed on codes of extracted candidates only selected in coarse search from signature memory. }
    \vspace{-15pt}
    \label{fig:acronym_algorithm_overview}
\end{figure*}
\textbf{Mapping Challenges and Parasitic Impacts.} Despite their theoretical efficiency, mapping large dimensional vectors (e.g., 128 to 768 dimensions) onto physical CAM array presents severe reliability challenges. High-dimensional vectors require long ML. These long interconnects introduce significant wire parasitic (RC delays) and severe IR drops along the array resulting in search accuracy drop\cite{11278694}. Additionally, these parasitic—coupled with the inherent device-to-device conductance variations of non volatile memory candidates distort the analog ML discharge rates. This distortion degrades the sensing margin between similar and dissimilar vectors\cite{11278694}. This directly impacts the recall rate of the search engine. Addressing these physical limitations high value resistor assisted CAM design has been addressed recently\cite{11278694}. However, the approach mitigates the problem to some extent and still dimension is expected to stay within 64-256 for best performance. Recent work \cite{kar2025ferroelectric} proposes dedicated popcounter block for each ML for higher accuracy for classfication applications where number of candidates are very limited within 100. This approach is not scalable for ANNS workload due to large overhead as this scales with dataset size. Hence discharge rate based Hamming distance representation is preferred. While large vector width is crucial for high recall, mapping them on hardware introduces inaccurate distance representation resulting in poor performance. To this end, ACRONYM introduces two stage search algorithm with limited vector width at each stage eventually function as twice size of vector width for final top-k selection. 
\\
\textbf{CAMs with Emerging Non-Volatile Memories.} While highly effective, traditional CMOS-based CAMs rely on bulky SRAM cells often requiring 10T transistors per cell for binary CAMs, which limits integration density for large scale applications like ANNS and suffers from high static leakage power. To overcome the area and power limitations of CMOS based CAM, non-volatile memory technologies has been explored for CAM implementation\cite{ni2019ferroelectric}. The emerging devices includes Resistive RAM (RRAM), Phase Change Memory (PCM), and Spin-Orbit Torque MRAM (SOT-MRAM). They offer zero standby leakage, back-end-of-line (BEOL) compatibility, and superior cell density\cite{haensch2023compute, roy2025breaking} compared to CMOS. Table \ref{tab:memory_comparison} presents a comparison of existing emerging non-volatile memory (NVM) candidates. While each technology offers distinct advantages and trade-offs, applications like ANNS demands stringent device-level requirements. Specifically, because dynamic ANNS will often involve frequent state updates, the underlying memory device must exhibit high endurance to ensure long-term reliability. Additionally, CAM cell area is crucial for scaling ANNS where millions of vectors need to be stored. Furthermore, a low operating voltage is strictly required to facilitate seamless monolithic integration with standard CMOS logic near the memory arrays. Finally, low search energy is critical for the ANNS applications. 

Given the necessity of data-distribution agnostic hashing based approach to address dynamic vector database search challenge and at the same time met the recall, latency and memory tradeoff. We need careful co-design of algorithm, architecture and circuits which is the goal of this work as referred at the end of introduction section.

\section{ACRONYM: Algorithm}


Fig.\ref{fig:acronym_algorithm_overview} illustrates the ACRONYM algorithmic overview. The whole pipeline can be described in two distinct phases: \circled{1} Offline setup and \circled{2} Online search.




\subsection{Offline Setup}

\textbf{\circled{1} Encoding.} The foundation of ACRONYM relies on transforming high-precision, floating-point vector data into a dense binary code. The binary code elements are independent and identically distributed (i.i.d.) variable and offers bit-level parallelism enabling efficient hardware execution. For encoding, we use random projection with sign-based binarization. In this method, a random projection matrix is generated and preserved for use during online search. The encoding of base vectors into binary code can be formulated as follows:
\begin{equation}
    \mathbf{C} = \text{sign}(\mathbf{H\times} \mathbf{B})
\end{equation}

where $\mathbf{H} \in \mathbb{R}^{D \times d}$ is the random projection matrix, with $D$ representing the encoded dimension and $d$ denoting the native vector dimension. The matrix $\mathbf{B} \in \mathbb{R}^{d \times N}$ represents the input batch of $N$ native vectors, and $\mathbf{C} \in \{0, 1\}^{D \times N}$ is the resulting encoded binary matrix. ACRONYM uses binary values 1 and -1 only in the projection matrix, $\mathbf{H}$. The information encoding capacity of such approaches has been already demonstrated\cite{achlioptas2003database, zhao2025hddb}. This way, ACRONYM enables efficient hardware execution of encoding without affecting performance.

\textbf{\circled{2} Partitioning.} For a particular dataset, all base vectors are projected and binarized into the high-dimensional binary space. Then the binary codes of full base vectors are partitioned into two distinct, contiguous blocks: coarse block and refinement block. The width of codes to be used in coarse block is dictated by parameter p in Fig.\ref{fig:acronym_algorithm_overview} and depends on the dataset size and the original vector width because they determine the information required for standard search performance\cite{thomas2021theoretical}. The code corresponding to the coarse block is loaded directly into the coarse CAM unit. This coarse block data remain strictly stationary during all online search where queries are streaming continuously, removing the massive data movement costs that fails traditional hardware architectures to run such algorithm. Another partition of binary base vectors is loaded into the refinement memory block. This memory also remains static during online search. After encoding, the full base vectors in original format are discarded. This reduces the memory required for the search to only code storage. Following encoding and updating the CAM and refinement memory block, the system is ready for online search.

\subsection{Online Search Pipeline}

When search is issued against a query, the system follows a deterministic two-stage pipeline to extract the nearest neighbors:

\textbf{\circled{1} Query Encoding.} The incoming search query is projected and binarized using the exact same random projection encoding scheme that was generated during the offline setup phase. The resulting binary code of the query is then split into two parts for coarse search and refinement, mirroring the base data. 

\textbf{\circled{2} Coarse Search.} The code for coarse search of the query is loaded into the query register of the coarse search CAM unit. Because the CAM unit computes distances in memory, it performs fully parallel distance computation against all base candidates' code, simultaneously. Ideally, this process executes in constant time, irrespective of the dataset size, and generates indices of the pool of coarse search candidates. Note that in our CAM implementation, this process does not move any data for computation and is thus fully in memory.

\textbf{\circled{3} Refinement Code Collection.} To avoid moving massive amounts of raw data and expensive distance computation to find final top-k candidates, ACRONYM reduces data movement by accessing only the refinement code corresponding to the pool candidates selected in coarse search. To realize these bandwidth savings at the system level, we utilize a memory-mapped fetch mechanism for efficient data transfer between refinement memory and refinement CAM block. 

\textbf{\circled{4} Final Refinement.} The refinement is performed against the refinement code of the query using the refinement CAM block. The distance computation is fully parallel and in memory like coarse search. This process generates the final pool of approximate top-k. Note that, ACRONYM uses information of $p$ bit for coarse and $D-p$ bit for refinement. This reduces requirement of single large wordline in CAM array. Thus meets the requirement of large hashing code and accurate CAM array operation for competent recall.

\begin{figure}
    \centering
    \includegraphics[width=0.9\linewidth]{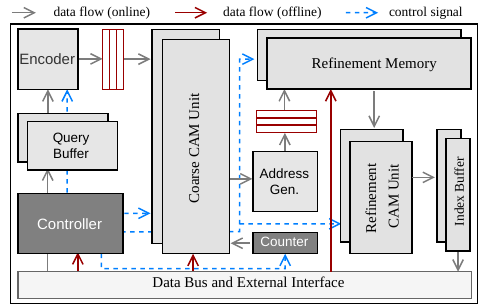}
    \vspace{-10pt}
    \caption{ACRONYM hardware architecture.}
    \vspace{-15pt}
    \label{fig:acronym_hardwware_arch}
\end{figure}

\section{ACRONYM: Hardware Architecture}
Fig.\ref{fig:acronym_hardwware_arch} illustrates an architectural overview of ACRONYM. The hardware operates in a fully digital domain. Data bus and external interfaces work as front end of the architecture which is used for offline configuration of the system and data loading into the memories (red arrows show the data flow direction). During search, queries are loaded into the query buffer and fed to encoder batch-wise for efficient processing that leverages matrix multiplication based formulation and uses novel XAC PE based systolic array. To ensure functionality in various clock domains in Encoder and Coarse CAM unit, a FIFO buffer is used to better pipeline the process. Coarse CAM Unit along with counter and address generator produce compact one-to-one memory mapped address of the pool candidates. To handle backpressure from the comparatively slow memory read, we use another FIFO buffer before the memory. Once the refinement codes of the candidates selected in coarse search is received and written into refinement CAM unit, it generates address of the top-k items in the form of multi hot encoding which are temporarily stored in index buffer before broadcasting to output bus. 
\begin{figure}
    \centering
    \includegraphics[width=0.8\linewidth]{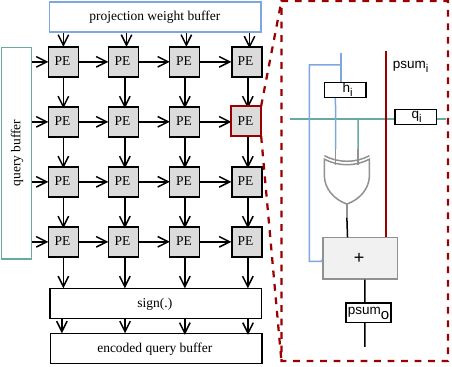}
        \vspace{-10pt}
    \caption{Encoder based on systolic array with XAC PE.}
    \label{fig:binEncoder}
    \vspace{-15pt}
\end{figure}

\subsection{Encoder}

Fig.~\ref{fig:binEncoder} illustrates the ACRONYM encoder designed for batch-wise query encoding. The encoding is formulated as a matrix multiplication followed by a binarization step. To avoid the use of expensive multipliers, we replace conventional multiply--accumulate (MAC) units with lightweight XOR and accumulation operation. The projection matrix weights are mapped from $\{1,-1\}$ to $\{0,1\}$ respectively. If the weight is $1$, XOR with $0$ leaves the query bit unchanged. If the weight is $-1$, XOR with $1$ produces the 1's complement of the query input. The weight bit is also fed into the carry-in of the adder to generate the 2's complement, effectively performing subtraction. Note that to perform the sign operation we placed the block at the streaming output end, i.e. bottom row. Before partial sums are stored they are binarized using a very simple sign functional module.

We use weight-stationary dataflow which is suitable for efficient execution of ACRONYM algorithm since the projection matrix remains stationary during the search process. The projection weights are preloaded into the systolic array in a $[d \times D]$ layout. During operation, the query batch is transposed and mirrored into a $[d \times b]$ format before being streamed through the systolic array. Queries are continuously fed during search and stored temporarily in the buffer, and during batch-wise encoding, queries are streamed into the encoder array. 

The encoded query vectors are produced row-wise in a staggered fashion and collected in the encoded query buffer, followed by the binarization stage. Thus, ACRONYM ensures encoded query vectors can be forwarded sequentially to the coarse search CAM unit. By eliminating multipliers and relying on XOR-based computation, the encoder becomes significantly faster and lightweight, enabling high-throughput encoding with high area and energy efficiency.

\begin{figure*}
    \centering
        \vspace{-5pt}
    \includegraphics[width=\textwidth]{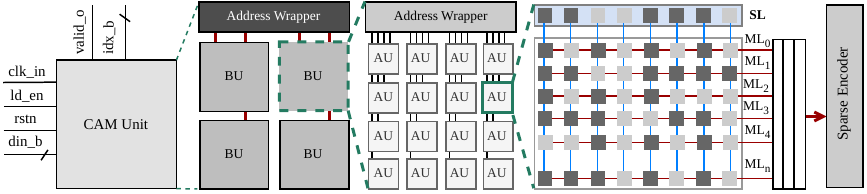}
        \vspace{-20pt}
    \caption{CAM unit's hierarchical organization to support efficient address encoding and stall free search during write. During top-k selection, items are latched inside AU where sparse encoder generates index of the 1s. Address wrapper in AU and BU adds header to locate them in the module.}
    \label{fig:cam_top_k_module}
        \vspace{-15pt}
\end{figure*}
\subsection{CAM Unit}
Fig.\ref{fig:cam_top_k_module} shows the detailed organization. It supports efficient address generation of pool candidates and fast memory access. The module is split into Bank Unit (BU) of CAM Array Unit (AU). AU is equipped with array followed by sparse encoder that generates 7 bit index of each 1 in input. Address wrapper in AU and BU add headers to represent location of the AU and BU in the module in a hierarchical order. CAM array of AU is primarily responsible for distance computation and ML outputs are used for top-k search. 
\subsubsection{\textbf{Distance computation}} 
\begin{figure}[b]
    \centering
    \vspace{-15pt}
    \includegraphics[width=0.8\linewidth]{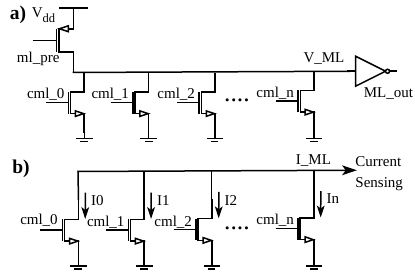}
    \vspace{-10pt}
    \caption{Distance representation scheme in CAM.}
    \vspace{-15pt}
    \label{fig:sensing_hamm_cam}
\end{figure}

To calculate the distance between query vector and a base vector which are binary we parallelize in bit level across CAM cell connected in single ML. Conventionally, each cell generates bit wise XOR output in form of high voltage drop at driver NMOS during mismatch and vice versa. Now similar to popcount we require count mechanism to sum the number of mismatches which is the Hamming distance between query and the base vector. Two primary sensing schemes are illustrated in Fig.~\ref{fig:sensing_hamm_cam}. 
In Fig.~\ref{fig:sensing_hamm_cam}(a), the ML is pre-charged prior to applying the search voltage to the search lines. For every mismatch bit, a discharge path is activated. As the number of mismatches increases, the discharge rate of the ML correspondingly increases. The time required for the ML voltage to drop below the switching threshold (clipping voltage) of the sensing inverter therefore depends on the Hamming distance. Once the threshold is crossed, the inverter generates a high signal at the ML\_out node. This approach offers low energy consumption and minimal sensing overhead, as it requires only a single inverter per ML. Current-based sensing as illustrated in Fig.~\ref{fig:sensing_hamm_cam}(b) offers finer precision but incurs higher area and energy due to voltage stabilizers and ADCs

 
 In the CAM array, distance computations are parallelized across all rows. To locate the rows storing the closest vectors to the query, ACRONYM uses the scheme illustrated in Fig.\ref{fig:sensing_hamm_cam}(a). In this approach, each ML is initially pre-charged, after which the encoded query is applied across the search lines (SLs). The discharge rate of the ML depends on the number of matches between the stored data and the query. Consequently, the ML voltage drops more slowly for larger Hamming distances and faster for smaller ones. Each ML is connected to an inverter that acts as a sense amplifier. When the ML voltage falls below Vdd/2, the inverter output switches to high. As a result, the output is a digital signal that initially starts at zero and transitions to high at different times depending on the Hamming distance. Note that, this approach does not require any ADC as output is digital and also for sensing we require a simple inverter compared to an expensive and power consuming sense amp. Additionally, it is quite energy efficient  due to ML discharge based technique compared to current and voltage-based approach where continuous power supply and large current are used.
\begin{figure}[b]
    \vspace{-5pt}
    \centering
    \includegraphics[width=0.8\linewidth]{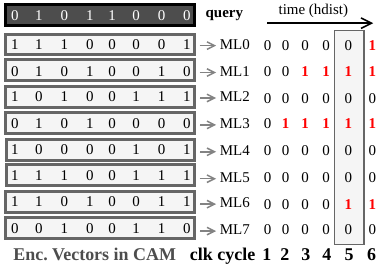}
    \vspace{-10pt}
    \caption{Time multiplexed distance representation. MLs with lower distance gets to 1 earlier than the higher distance.}
    \vspace{-10pt}
    \label{fig:ml_distance_timePlex}
\end{figure}
\subsubsection{\textbf{Top-k Selection}} ACRONYM algorithm finds a pool of candidates in coarse search instead of top-1 item. In most efficient algorithms, Quickselect in this case, the time complexity is O(N) where N is the number of items against the search is taking place. This means linear delay with the increment of pool size. In hardware this is expensive too due to large number of comparator blocks and thus pose challenges to use in realistic ANNS with millions of items. The proposed CAM array based design generates multi hot encoded ML output where position of 1 represent item's index. Fig.\ref{fig:ml_distance_timePlex} depicts the mechanism of distance representation where MLs with lower distance gets to 1 faster than the higher distance. Now the goal is extract the pool with most similar items of certain size. One approach is at every cycle, we perform a global population count to determine the number of 1s across the ML outputs. Once the target pool size is reached, the corresponding outputs are latched. This approach eliminates the need for expensive ADCs for top-k selection. However, it still needs an expensive popcount block with tree like reduction results in large area and power consumption. To this end, we use time based latching where we latch the output of the CAM array after a pre-determined time based on offline calculated distance and corresponding time from the characterization of CAM array. We can change the time to vary the number of candidates to be used in the pool at any time. This approach generates pool size of approximate value and not a very fixed one. However, coarse search pool doesn't need to be of exact size in ANNS. In this implementation, we don't need any counter and register for intermediate data storage. The latched outputs form a multi-hot encoded vector, where the index of each 1 corresponds to the index of a selected candidate. This information can then be directly used to retrieve the corresponding codes from the refinement memory for pool refinement.




\subsection{Memory Mapping} ACRONYM algorithm stores different codes of identical items in coarse CAM and refinement memory. During coarse search, the CAM unit generates multi hot encoded signal from the array where the position of each bit 1 represents the index which needs to be accessed in the refinement memory for refinement code. Now the challenge is efficiently access the memory for refinement code of the pool items. To achieve this, ACRONYM introduces one to one mapping. The coarse CAM block is structured in hierarchical order enabling efficient address generation(Fig.\ref{fig:cam_top_k_module}). The array unit (AU) is the underlying unit responsible for CAM array search of shape 128x128. Bank unit (BU) is array of the AUs (shape $16\times16$). Finally the coarse module is made of the BUs arranged in a square array of $8\times8$. This configuration alone can support 2M items with 32MB of memory size. We ensure identical shape of coarse CAM unit with memory unit. For address generation, we start with the index of the 1s (7 bit for each 1s location in AU) generated by sparse encoder and then the AU address (8 bit/AU) included as header of all 1s in the AU and finally the index of the BU (6 bit/BU) is included as header of the AUs with 1s. This hierarchical address generation offers fast and easy address management. Besides, it offers a simpler address decoder design for the memory where the decoder decodes in parallel starting with the BU index, then AU index and finally the actual index of the 1 in the array. 

\begin{figure}
    \centering
    \vspace{5pt}
    \includegraphics[width=0.7\linewidth]{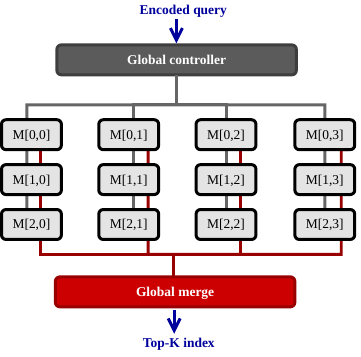}
    \vspace{-15pt}
    \caption{ACRONYM scale out to support large scale dataset.}
    \vspace{-20pt}
    \label{fig:scaling}
\end{figure}
\subsection{Dynamic Update Policy}

To support frequent vector insertions and deletions in dynamic databases, ACRONYM employs a parallel, hardware-managed update policy that avoids global stalls and minimizes latency.

\textbf{Batching and Parallel Bookkeeping.}
The central controller maintains a \emph{Data Insertion Queue} to decouple updates from query execution. Updates are processed in batches to amortize non-volatile memory (NVM) write latency. Bookkeeping metadata are handled off the critical path, ensuring uninterrupted distance computation.

\textbf{Bank-Interleaved Writes for Stall-Free Search.}
ACRONYM exploits the hierarchical CAM organization BUs and AUs to enable fine-grained updates. During insertion, only a single AU is locked for writing, while remaining AUs continue serving queries. Due to uniform distribution from random projection encoding and large coarse search pool size, temporarily masking one AU introduces negligible recall degradation while preserving throughput and bounded tail latency. A priority-based allocation policy directs writes first to empty AUs, and otherwise to AUs with the highest number of deletions dictated by deletion queue. As deleted spots are filled up through insertions, the index is cleared from the queue.

\textbf{Handling Write Hotspots.}
To prevent write hotspots and limit system reliability, the controller periodically remaps logical-to-physical AU assignments. This mechanism distributes write stress uniformly, ensuring long-term NVM reliability.

\subsection{Multi-Module Architecture for Scaling}
To scale ACRONYM beyond a single CAM module's capacity limited by single chip area constraint, we demonstrate scale out with CAM as modular unit that enables collectively indexing of large applications with hundreds of millions of items. Fig.\ref{fig:scaling} illustrates the scale out architecture where a global controller broadcasts query and a single latch time threshold simultaneously to all CAM modules, which then execute coarse search and local refinement in fully parallel — with no inter-module communication for search. A single latch threshold suffices across all partitions because the scale out approach is simply intended for capacity scaling. Each module returns a local top-k candidate pool encoded as a multi-hot index vector. The global merge simply merges these local pools to produce the final result. 

\begin{figure*}
    \centering
    \vspace{-15pt}
    \includegraphics[width=\textwidth]{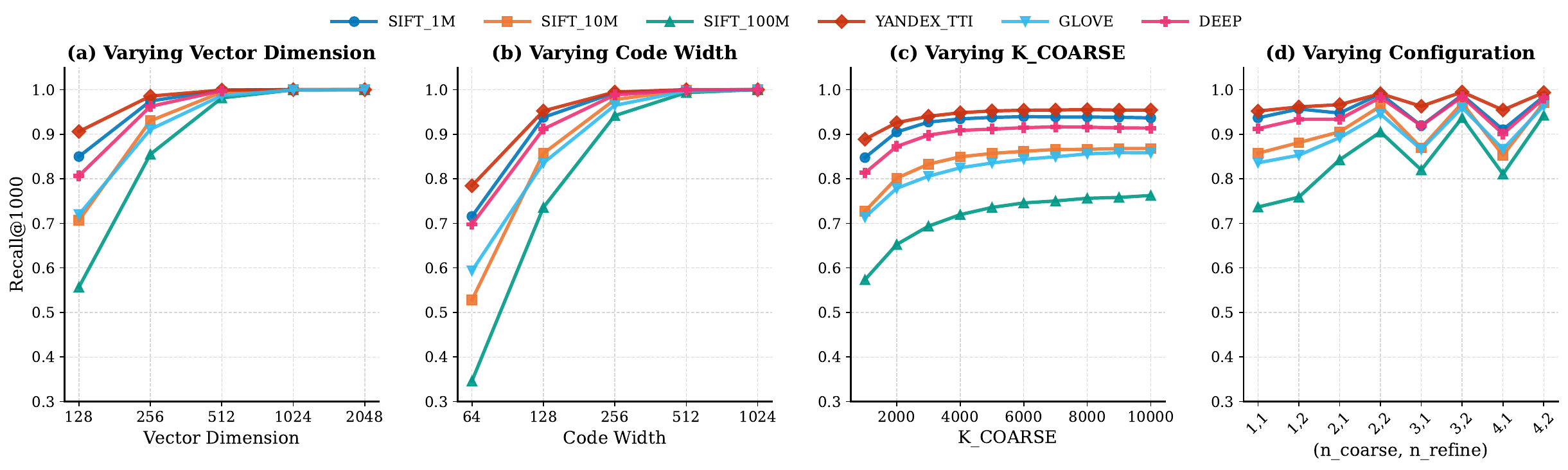}
    \vspace{-25pt}
    \caption{a) Search using full vector at larger dimension has higher recall. But larger dimension require larger memory footprint and increased error due to parasitic of CAM array. b) ACRONYM two stage search depicts high recall at lower code width. c) Performance at varying pool size at fixed code width of 128. This increases recall by increasing pool size generated from coarse search. d) Parallel coarse search with smaller code width enables accurate hardware mapping with high recall. }
    \vspace{-10pt}
    \label{fig:recall}
\end{figure*}

\begin{figure}[b]
    \centering
    \vspace{-10pt}
    \includegraphics[width=0.8\linewidth]{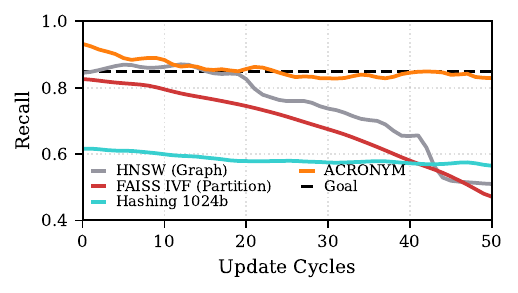}
    \vspace{-15pt}
    \caption{ACRONYM Recall during frequent updates.}
    \vspace{-15pt}
    \label{fig:recall_vs_update_comp}
\end{figure}

\begin{table}[htbp]
\vspace{-10pt}
\centering
\caption{Datasets used for ACRONYM evaluation}
\vspace{-10pt}
\label{tab:datasets}
\begin{tabular}{lccc}
\toprule
\textbf{Dataset} & \textbf{Dimensions} & \textbf{Size} & \textbf{Distance} \\
\midrule
GloVe\cite{pennington2014glove}       & 100 & 1,183,514   & Inner Product        \\
DEEP\cite{babenko2016efficient}        & 96  & 9,990,000   & Euclidean      \\
Yandex TTI\cite{bigann_dataset}  & 200 & 9,990,000   & Inner Product  \\
SIFT1M\cite{jegou2011searching}       & 128 & 1,000,000 & Euclidean      \\
SIFT10M\cite{jegou2011searching}       & 128 & 10,000,000 & Euclidean      \\
SIFT100M\cite{jegou2011searching}       & 128 & 100,000,000 & Euclidean      \\
\bottomrule
 \vspace{-25pt}
\end{tabular}
\end{table}

 \section{Experiments and Results}
\label{Experiments and Results}

\textbf{Datasets.} Table \ref{tab:datasets} summarizes details on the datasets used for evaluating and benchmarking ACRONYM’s algorithmic and hardware design choices. \\
\textbf{Baseline Algorithms Implementations.} We use graph based, partition based and quantization based algorithms as baseline. For graph based approach, HNSW and DiskANN implemented in python have been used. For partition based we use standard FAISS-IVF and for partition with quantization approach we use FAISS-IVFPQ from FAISS library\cite{douze2025faiss} implemented in python. For CPU and GPU execution, we use a dual-socket Intel Xeon Gold 6548Y+ system with 64 physical cores (128 hardware threads) operating at up to 4.1 GHz. The system is equipped with 880 GB of main memory distributed across two NUMA nodes. For GPU-based baselines, we use two NVIDIA L40S GPUs, each with 46 GB of device memory, running CUDA 13.2. During profiling in GPU, we used warm up and multiple run to ensure accurate representation. For latency measurement, we use built in $perf\_counter$. \\
\textbf{ACRONYM Algorithm Evaluation Methodology.}
For algorithmic performance evaluation, we implemented the algorithm in python. We have used FAISS library for Hamming distance search that utilizes CPU supported popcount operations.  The inefficiency of CPU and GPU execution of ACRONYM is presented in Fig.\ref{fig:profling_acronym}. \\
\textbf{ACRONYM Hardware Implementation.} For functional validation, we have designed encoder block, address generation block, memory decoder block and controller in RTL. We performed functional verification of each module individually. For latency, energy and area evaluations we performed synthesis in TSMC 65nm GP with 200MHz for the modules. We used Synopsys VCS for functional verification and Synopsys Design Compiler for synthesis. For refinement memory, we used HBM with bandwidth of $1280 GB/s$ with write energy $4pJ/b$\cite{10454440}. For CAM units, we used data from the array level characterization data of various CAM implementations available in recent literature and HSPICE simulations as presented in Table.\ref{tab:memory_comparison}. Based on the synthesis results and CAM array characterization data, we have developed a model to estimate end-to-end latency, energy and area of the overall system. We have used CAM arrays of shape $128\times128$. For encoder, we have used systolic array block of size $64\times64$.

\subsection{Results: Algorithm Validation}

To validate the effectiveness of information representation in binary form and the end to end ACRONYM algorithm, we performed experiments and demonstrate the correlation in Fig.~\ref{fig:recall}.
\subsubsection{\textbf{Recall Vs Vector Width.}}
To quantify the impact of binary encoded vector width on recall, we measured recall for different vector dimension. To ensure sustained performance across a variety of datasets, we used four distinct datasets of different native dimensions and sizes. Fig.\ref{fig:recall}(a) depicts the full vector search recall vs vector dimension, where we simply select top-k candidates based on the full encoded vector. For large dimensions recall increases as more information is encoded and this shows the information encoding capacity in the binary encoded vectors. However, a larger vector dimension introduces two challenges. \circled{1} it requires large memory to store and pose challenges to scale for large datasets with millions to hundreds of millions of items \circled{2}  executing the search using in-memory schemes such as CAM based approaches gets inaccurate. This is because of the increase of the number of cells connected to a single ML  which results in considerable leakage current passing through the mismatch cells and the excessive parasitic RC components. ACRONYM algorithm overcome the issue. Fig.\ref{fig:recall}(b) shows ACRONYM algorithm's recall at varying signature width. Note that the proposed algorithm improves over full vector search at the same signature width thanks to the two stage search even though ACRONYM uses a lower dimension vector. Thus, ACRONYM not only solves the memory mapping issue on CAM arrays at reduced storage, but improves over full vector search baseline. 
\begin{figure}[b]
    \centering
    \vspace{-15pt}
    \includegraphics[width=\linewidth]{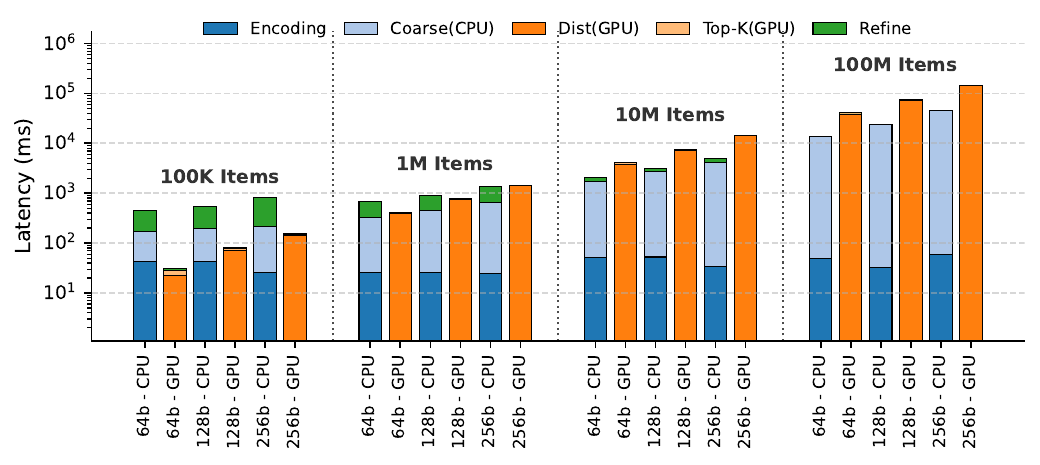}
    \vspace{-20pt}
    \caption{Profiling ACRONYM algorithm in CPU and GPU. Latency in logscale shows distance computation dominates the latency both in CPU and GPU.}
    \vspace{-10pt}
    \label{fig:profling_acronym}
\end{figure}

\subsubsection{\textbf{Recall Sensitivity.}} ACRONYM uses time-multiplexed approximate top-k selection to avoid any hardware overhead. This approach does not ensure exact top-k candidate selection; instead, it selects approximate top-k candidates. We conducted an experiment to measure the sensitivity of recall to the pool size i.e., $K\_COARSE$. Fig.\ref{fig:recall}(c) shows a larger pool size increases the recall but it saturates soon. Waiting for longer time for large candidate pool size improves the recall but at the cost of increased memory access time for refinement code of inflated pool. 

\subsubsection{\textbf{ACRONYM Recall Boosting.}} To further improve recall, ACRONYM offers a few potential configurations. During coarse search, we can use multiple coarse blocks or refinement blocks. This approach will increase the memory overhead but still offers realistic mapping CAM arrays with 64 to 256 columns. This type of configuration is particularly valuable for ANNS applications with large native vector dimensions or large search space where small code widths often suffer to achieve high recall. Fig.\ref{fig:recall}(d) demonstrates recall at a fixed $K\_COARSE$ of 5000 for various configurations. Note that the coarse block increment has a higher impact on recall improvement than refine block as coarse block is responsible for a more exhaustive and selective pool generation. 
\subsubsection{\textbf{Recall Robustness in Dynamic DB}} To highlight robustness during dynamic DB update, we used a 1M size dataset and at each cycle we delete 100K ($10\%$) and add 100K randomly generated vectors. We also used newly generated queries from random distribution. We measured ground truth through an exact search at each update cycle and measured average recall on 1000 queries. Fig.\ref{fig:recall_vs_update_comp} illustrates recall vs update cycle where ACRONYM recall is not degrading and demonstrates robustness during update with out- of-order distribution. The initial degradation is due to data distribution shifts from initial SIFT vectors to random vectors and difference exists in different cycle's data and query.

\begin{figure}
    \centering
        \vspace{-10pt}
    \includegraphics[width=0.8\linewidth]{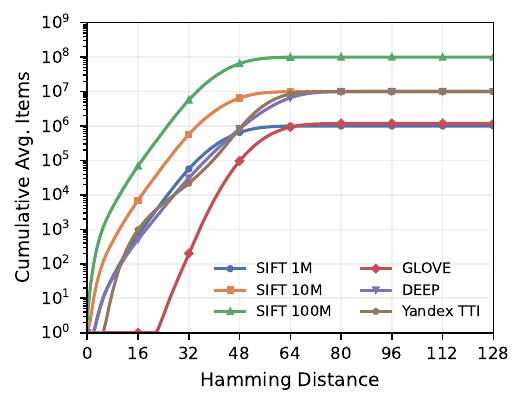}
    \vspace{-15pt}
    \caption{Average cumulative distribution of base item based on Hamming distances during search. This validates the ACRONYM's top-k selection approach where latch at a fixed cycle will generates approximated pool size.}
    \vspace{-15pt}
    \label{fig:cum_dist_pool_hdist}
\end{figure}

\vspace{-5pt}
\subsection{Profiling ACRONYM Algorithm}
We profiled the ACRONYM algorithm on CPU and GPU architectures to highlight their limitations in processing ACRONYM algorithm and to motivate the use of in-memory computing approaches such as CAM. In CPU, for Hamming distance computation we use FAISS\cite{douze2025faiss} \texttt{IndexBinaryFlat}. As distance computation and top-k search are both fused in the implementation we report them together. For GPU, we use PyTorch based implementation with batched queries and tiled database scans for distance computation and top-k search. We used 128 dimension base vectors of different scale e.g., $N \in \{10^5, \dots, 10^8\}$ and  We sweep the binary code width $W \in \{64, 128, 256\}$ bits where each stage uses $W$ bits and packed in bytes for efficient fetching from memory. Fig\ref{fig:profling_acronym} illustrates the result which shows in CPU the coarse search cost scales with $N$ and dominates the overall computation time. GPU shows speedup over CPU at small size dataset as the GPU cache in lower hierarchy close to compute unit can accommodate the  entire encoded base vectors thus saves the data movement cost. But it quickly gets out of available cache and requires multiple hop-on and hop-off for base vectors through GPU memory hierarchy for each query. This adds on top of the increased distance computation cost. At 10M scale, GPU lags behind CPU as memory transactions become very prevalent. Ultimately, at large $N$, both CPU and GPU execution times are heavily dominated by distance-oriented, memory-bound kernels. Thus, to achieve realistic throughput from ACRONYM algorithm, we require specialized hardware that naturally offers superior distance computation with reduced data movement. Given the context, CAM offers static in-memory distance computation and eliminates the heavy data movement.
\begin{figure}
    \centering
    \vspace{-10pt}
    \includegraphics[width=1.1\linewidth]{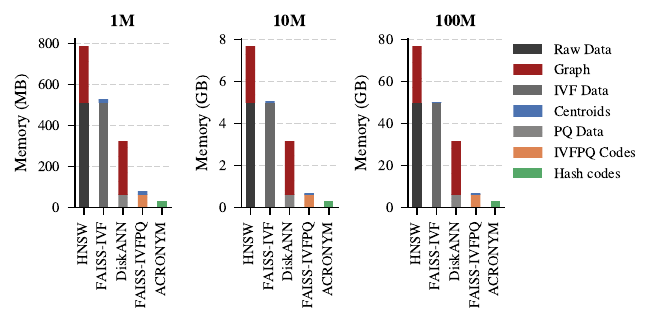}
    \vspace{-25pt}
    \caption{Memory footprint comparison with existing approaches for different data size. }
    \vspace{-15pt}
    \label{fig:memory_compression}
\end{figure}

\begin{figure}[b]
    \centering
    \vspace{-10pt}
    \includegraphics[width=0.8\linewidth]{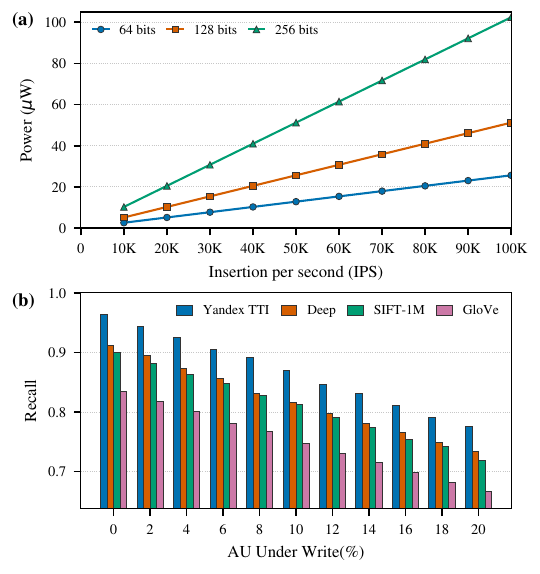}
    \vspace{-15pt}
    \caption{ a) Memory cell write power during update, b) Recall during ongoing updates where AU under write is excluded from search. }
    \vspace{-10pt}
    \label{fig:dynamic_update_cost}
\end{figure}

\begin{figure*}
    \centering
    \vspace{-15pt}
    \includegraphics[width=\textwidth]{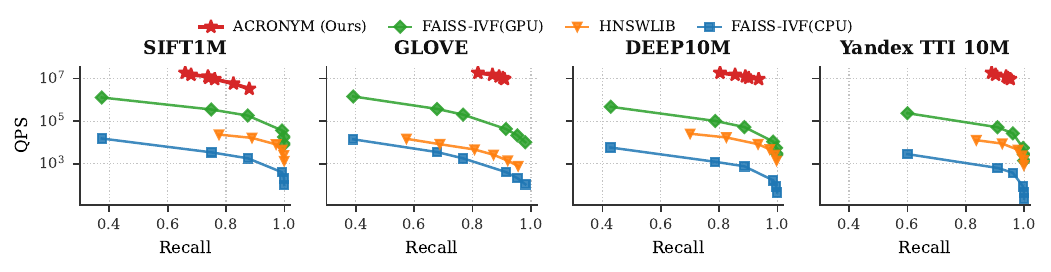}
    \vspace{-30pt}
    \caption{Throughput vs performance tradeoff comparison.}
    \vspace{-10pt}
    \label{fig:qps_vs_recall_comp}
\end{figure*}

\begin{figure}[b]
    \centering
    \vspace{-10pt}
    \includegraphics[width=0.8\linewidth]{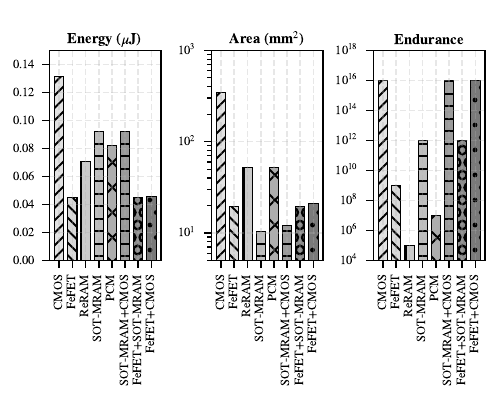}
    \vspace{-25pt}
    \caption{Comparison of different CAM choices in terms of system perspective: energy per query, cell area, system endurance. Lower energy and cell area at higher endurance is better.}
    \vspace{-20pt }
    \label{fig:dsa_cam}
\end{figure}

\subsection{Results: Hardware Validation}
\subsubsection{\textbf{Time Multiplexed Top-k Selection.}}
ACRONYM proposed time-multiplexed top-k selection quality is dependent on distribution of items across Hamming distances during the search operation. To show the effectiveness of the method, we measured and plotted the distribution in the form of cumulative distribution with Hamming distance because in the proposed time-based method, the top-k pool size directly depends on the time elapsed during the discharge phase because the Hamming distance determines the discharge time. Fig.\ref{fig:cum_dist_pool_hdist} shows the average cumulative distribution for various datasets measured from 1000 queries. The illustration shows that most of the candidates are distributed in lower Hamming distance region but shows gradual increase. This dictates items are distributed across Hamming distance in a window which is important for top-k selection. 
\subsubsection{\textbf{Memory Footprint.}}
ACRONYM does not require raw vector storage like HNSW and FAISS-IVF. Instead, it generates binary codes of dimensions 8-32 Byte per vector and stores them in coarse CAM block and refinement memory. This allows ACRONYM to be memory efficient. Fig.\ref{fig:memory_compression} illustrates the memory breakdown and compression ratio of various methods for various dataset sizes. HNSW requires the storage of the graph structure and raw data. Diskann stores quantized data with the associated graph. FAISS-IVF require the storage of centroids and raw data, while FAISS-IVFPQ stores quantized data with centroids and codebook. ACRONYM stores only binary codes partitioned in coarse memory and refinement memory.

\subsubsection{\textbf{Dynamic Update.}} Although ACRONYM supports dynamic updates without interrupting search, dynamic updates introduce power spikes due to random write operations in the memory cells. Fig.\ref{fig:dynamic_update_cost}(a) shows the power required to write data in the cells vs. insertion per second (IPS). The plot shows power increases linearly with the increase of IPS. Apart from power, dynamic update support requires the AU blocks under write to be isolated from ongoing search and cause the active items inside those AUs not to be searched. Thus, if the number of AU blocks under write increases too much, it might blind the system and hence will causes recall drop due to reduced search space. To show the impact, we varied the percentage of blocks under write and then omit items belongs to the blocks during search and calculated the recall. Fig.\ref{fig:dynamic_update_cost}(b) illustrates the impact for varying the fraction of the blocks that are busy from  0 to $20\%$. The results inspire continuous but small chunk size data insertion instead of bulk insertion which will reduce power and ensures high recall by not locking many active item blocks.

\subsubsection{\textbf{Throughput Vs Recall.}}
To benchmark ACRONYM with CPU and GPU-powered baseline i.e., graph-based HNSW and partition-based FAISS-IVF, we performed queries per second (QPS) vs recall evaluation as illustrated in Fig.\ref{fig:qps_vs_recall_comp}. FAISS library provides support for execution on both CPU and GPU. Graph-based HNSW is CPU-friendly; hence, the plot shows only the CPU-based results. The illustration shows that ACRONYM offers higher throughput because of reduced and efficient data movement. The main bottleneck of ACRONYM is data movement between memory unit and CAM refinement block. However, ACRONYM efficient one-to-one memory mapping and usage of emerging memory with fast read can reduce the throughput.

\vspace{-10pt}
\subsection{Design Space Exploration: CAM}
ACRONYM architecture and hardware implementation are orthogonal to any CAM technology. This offers design space exploration opportunity to achieve the best tradeoff. Note that, in the proposed architecture, the overall system search energy and area is dominated by the Coarse CAM block. On the other hand, refinement CAM block has negligible energy and area footprint due to its small size. However, this block is very write-heavy and thus require high endurance for long term reliability. In this scenario, CAM device with low endurance are not suitable for long term reliability if used in refinement block. However, they can be used in coarse block since that does not demand high endurance as item vectors remain there and updates relatively less frequently. We have used SOTA CAM design available for all the CAM technology as summarized in Table.\ref{tab:memory_comparison} and showed the energy, area and system endurance comparison for various design choices for a 1M size ANNS application as presented in Fig.~\ref{fig:dsa_cam}. It depicts that FeFET and SOT-MRAM offer good tradeoff in terms of area and energy. However, pure FeFET-based implementation suffers from low endurance and thus limits the overall life cycle of the system. On the other hand, SOT-MRAM based implementation offers better yet limited life cycle given the long operational period of ANNS system. The hybrid approach offers the best tradeoff where FeFET as coarse block and CMOS as refine block delivers lower energy and area at the highest endurance. This choice will be used to benchmark ACRONYM hardware afterwards. 
\vspace{-10pt}
\subsection{Variation Analysis}
\begin{figure}[t]
    \centering
    \vspace{-0pt}
    \includegraphics[width=\linewidth]{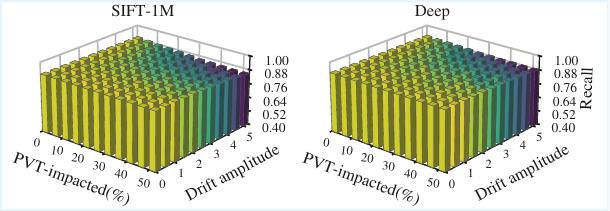}
    \vspace{-20pt}
    \caption{Search performance under non-ideality due to impacts on proposed approximated top-k selection.}
    \vspace{-5pt}
    \label{fig:pvt_impact}
\end{figure}

ACRONYM introduces approximate top k selection which depends on the relationship between discharge time and Hamming distance. This is highly dependent on device characteristics, latch time sensitivity and process, voltage, time (PVT) variability. Device level Discharge time vs Hamming distance relationship is known and can be stabilized further with reference resistor and thus can be considered accordingly\cite{11278694}. Second, latching time fluctuation is rare at $ns$ range as digital clocks are accurate enough at this scale. Third point of vulnerability, PVT variation is inevitable and thus may distort known discharge time vs Hamming distance relationship. To model the PVT variation impact on recall, we considered the percentage of array impacted and then to quantify the impact of variability on distance misrepresentation we use distance noise that we varied across 0 to 5. Fig.\ref{fig:pvt_impact} shows the impact of PVT variation on recall. The illustration depicts performance degradation with the increase of the array under PVT variation with high distance impact. However, the gradient of performance drop is low that shows system robustness against PVT variability.

\begin{table}[htb]
\centering
\vspace{-10pt}
\caption{Encoder Specifications}
\vspace{-10pt}
\label{tab:encoder_specs}
\begin{tabular}{@{}ll@{}}
\toprule
Process           & TSMC 65nm GP   \\
Frequency         & 200 MHz        \\
Precision (input) & INT16          \\
Array shape       & $64 \times 64$ \\
Throughput        & 1.638 TOPS     \\
Supply            & 0.9--1.1 V     \\
Area              & 2.91 mm$^2$    \\
Power             & 132.53 mW      \\ \bottomrule
\end{tabular}
\vspace{-20pt}
\end{table}
\subsection{Overhead Analysis}
\subsubsection{\textbf{ACRONYM Encoder.}} Functionally verified RTL design of the $64\times64$ shape array has been synthesized using TSMC 65nm PDK for the determination of the area, power and latency. We used $int16$ for base vector dimension and partial sum output. The array supports weight stationary (WS) dataflow only as the projection matrix remains unchanged and hence WS offers best reuse at reduced PE complexity. The characterization is summarized in Table~\ref{tab:encoder_specs}. Note that the area of the $64\times64$ array  made of XAC unit is $2.91mm^2$ where a $16\times16$ array of MAC unit at the same technology node takes a $3.02mm^2$ of area as reported in DianNano\cite{chen2014diannao}. This is because XAC avoids expensive multipliers and instead uses simple XOR units. We have limited the clock frequency at 200MHz to match the memory access throughput.

\subsubsection{\textbf{Memory overhead.}} In ACRONYM CAM is at the center of computing and acts as half of the storage. Although high density emerging memory based CAM has been introduced, the CAM memory array dominates total area of the system. Additionally, this limits the capacity of scaling of the algorithm for billion scale. For refinement memory, high bandwidth and density are the key metrics. Additionally non-volatility helps to reduce overall system energy consumption.

\begin{table}[t]
\vspace{-5pt}
\centering
\small
\setlength{\tabcolsep}{3pt}
\renewcommand{\arraystretch}{1.1}
\resizebox{\columnwidth}{!}{
\begin{tabular}{lccccc}
\toprule
\textbf{System} & \textbf{ANNA\cite{lee2022anna}} & \textbf{SeIM\cite{liu2025seim}} & \textbf{Proxima\cite{11433588}} & \textbf{FANNS\cite{yuan2025fanns}} & \textbf{ACRONYM} \\
\midrule
Platform & ASIC & NMA\footnotemark[1] & NMA\footnotemark[1] & FPGA & \textbf{IMA\footnotemark[2]} \\
Distribution-agnostic\footnotemark[3] & \ding{55} & \ding{55} & \ding{55} & \ding{55} & \textbf{\ding{51} (Fig.\ref{fig:recall_vs_update_comp})} \\
Algorithm & PQ+Partition & PQ+Partition & Graph & Partition & Hashing \\
Search during DB update & Stall & Stall & Stall & Stall & \textbf{Regular} \\
In Memory Distance & \textbf{\ding{55}} & \textbf{\ding{55}} & \textbf{\ding{55}} & \textbf{\ding{55}} & \textbf{\ding{51} (CAM)} \\
Separate Top-K Logic & \textbf{\ding{51}} & \textbf{\ding{51}} & \textbf{\ding{51}} & \textbf{\ding{51}} & \textbf{\ding{55} (time-latch)} \\
\bottomrule
\end{tabular}
}
\footnotesize{
\textsuperscript{1}Near-Memory Acceleration \quad
\textsuperscript{2}In-Memory Acceleration \quad
\textsuperscript{3}Stable recall under update from dynamic distribution \quad
}

\caption{Comparison with prior ANNS accelerator}
\vspace{-35pt}
\label{tab:anns_accelerator_comp}
\end{table}

\subsection{Prior ANNS Accelerators and ACRONYM}

Recent hardware acceleration efforts for ANNS explore diverse architectures to mitigate memory bandwidth bottlenecks and redundant computations in high-dimensional retrieval. ASIC designs such as ANNA \cite{lee2022anna} reduce data movement by operating on compressed vectors with parallel lookup pipelines, while FPGA-based FANNS \cite{yuan2025fanns} eliminates redundant distance computations via subspace partitioning. To address the large memory footprint of graph-based methods, recent work shifts toward near-memory acceleration: SeIM \cite{liu2025seim} offloads memory-bound operations to DRAM banks through a hierarchical design, and Proxima \cite{11433588} integrates compute units near 3D NAND arrays. However, these accelerators primarily target static databases, as they are built on graph-based and partition-based algorithms; consequently, performance degrades when serving dynamic databases. Table \ref{tab:anns_accelerator_comp} summarizes the comparison between existing ANNS accelerators and ACRONYM, highlighting that ACRONYM supports dynamic updates, performs distance computations entirely in memory, employs a time-latch-based top-k selection mechanism, and enables updates during search.
\section{Conclusion}

Dynamic vector databases pose a fundamental challenge for existing ANNS methods: graph-based and partition-based indices degrade rapidly under continuous insertions and deletions because their construction depends on an initial estimate of the data distribution, necessitating costly periodic reconstruction. ACRONYM addresses this by combining data-distribution-agnostic random projection encoding with a two-stage CAM-based in-memory search, eliminating index rebuilding entirely while achieving greater than $90\%$ recall at $8\times10^6$ queries per second within a $32 MB$ memory footprint and $2.56\mu J$ per query — outperforming CPU and GPU baselines across million-scale datasets under dynamic update workloads. The systolic-array XOR-and-Accumulate encoder avoids expensive multipliers, the time-multiplexed top-k selection eliminates ADCs and digital sorting logic, and the bank-interleaved update policy enables simultaneous search and insertion with negligible recall degradation. Design space exploration across CMOS, FeFET, ReRAM, SOT-MRAM, and PCM technologies identifies a hybrid FeFET-coarse plus CMOS-refinement configuration as the optimal balance of energy, area, and long-term endurance, providing a practical roadmap for deploying ACRONYM for real life applications.
\section{Acknowledgment}
This work was supported in part by CoCoSys and PRISM, two centers in JUMP 2.0, a Semiconductor Research Corporation (SRC) program sponsored by DARPA.

\bibliographystyle{ACM-Reference-Format}
\bibliography{acronym}

\end{document}